\documentclass[aps,prb,twocolumn,groupedaddress]{revtex4-1}
\usepackage{graphicx}  % needed for figures
\usepackage{dcolumn}   % needed for some tables
\usepackage{amssymb}   % for math
\usepackage{amsmath}   % for math
\usepackage{amsthm}   % for math
\usepackage{bm,bbold}        % for math
\usepackage{comment} % for comments
\usepackage{color}
\begin{document}

\title{Effective theory of vortices in two-dimensional spinless chiral $p$-wave superfluids}

\author{Daniel Ariad}
%\email[]{ariad@ymail.com}
\affiliation{Department of Physics, Ben-Gurion University of the Negev, Beer-Sheva~8410501, Israel}
\author{Babak Seradjeh}
\affiliation{Department of Physics, Indiana University, Bloomington, Indiana 47405, USA}
\author{Eytan Grosfeld}
\email[]{grosfeld@bgu.ac.il}
\affiliation{Department of Physics, Ben-Gurion University of the Negev, Beer-Sheva~8410501, Israel}

%\date{\today}

\begin{abstract}
We propose a $\mathbb{U}(1) \times \mathbb{Z}_2$ effective gauge theory for vortices in a $p_x+ip_y$ superfluid in two dimensions. The combined gauge transformation binds $\mathbb{U}(1)$ and $\mathbb{Z}_2$ defects so that the total transformation remains single-valued and manifestly preserves the the particle-hole symmetry of the action. The $\mathbb{Z}_2$ gauge field introduces a complete Chern-Simons term in addition to a partial one associated with the $\mathbb{U}(1)$ gauge field. The theory reproduces the known physics of vortex dynamics such as a Magnus force proportional to the superfluid density. More importantly, it predicts a universal Abelian phase, $\exp(i\pi/8)$, upon the exchange of two vortices. This phase is modified by non-universal corrections due to the partial Chern-Simon term, which are nevertheless screened in a charged superfluid at distances that are larger than the penetration depth.
\end{abstract}

%\pacs{}

\maketitle

\section{Introduction}
The two-dimensional spinless chiral $p$-wave superfluid is the minimal model for describing the properties of many realizations of topological superfluids and superconductors: topological insulator-superconductor interfaces~\cite{fu2008superconducting,sau2010generic,grosfeld2011observing}; the layered material Sr$_2$RuO$_4$ \cite{mackenzie2003superconductivity,grosfeld2011proposed,seradjeh2011unpaired}; some cold atom systems~\cite{gurarie2005quantum,grosfeld2007predicted}; and certain spin models admitting anyon excitations~\cite{kitaev2006anyons}. In this model,
the vortex defects of the phase of the pairing order parameter bind Majorana zero modes that endow them with non-Abelian exchange statistics~\cite{moore1991nonabelions,nayak1996quasihole,read2000paired,ivanov2001non-abelian}. Thus, they have been proposed as potential candidates for fault-tolerant, topological quantum information processing~\cite{freedman2003topological,kitaev2003fault,alicea2012new}. In addition, they are expected to admit a quantized Abelian exchange phase that plays an important role in proposals for {universal} topological quantum computation with vortices~\cite{bonderson2013twisted}. It is therefore quite important to formulate a cogent theory that accounts for the dynamics of vortices.

In previous work on this system, a low-energy effective action has been derived by the standard gradient expansion method~\cite{volovik1987analog,volovik1987peculiarities,goryo2000vortex,stone2004superfluidity,lutchyn2008gauge,roy2008collective}, shedding light on the collective response of the superfluid to external electromagnetic fields. However, in this derivation vortices have been generally left out. It appears then that the Abelian exchange phase of vortices, while surmised from the conformal properties of its edge states or the properties of candidate bulk wavefunctions~\cite{moore1991nonabelions,read2000paired,fendley2007edge}, has never been derived from a microscopic model~\cite{fradkin1998chern,hansson2013effective}. Consequently, it remains unclear whether bulk vortices in a chiral $p$-wave superfluid or superconductor exhibit this exchange phase and, if so, to what degree it is universal or how it is affected by the physics of the system.

To answer these questions, in this paper, we derive a $\mathbb{U}(1)\times \mathbb{Z}_2$ effective gauge theory that handles vortex defects properly. 
The $\mathbb{U}(1)$ gauge field is governed by an action that is identical to the one previously derived by gradient expansion, including a partial Chern-Simons (CS) term. 
Interestingly, a $\mathbb{Z}_2$ gauge field emerges in the effective theory governed by a new \emph{full} Abelian CS term. 
We show that the coefficient of the partial CS term is not a universal quantity and depends on the details of dispersion and higher-energy behavior of the system. The full CS term of the $\mathbb{Z}_2$ gauge field is, on the other hand, a truly topological term with a quantized coefficient. We calculate the exchange angle of two vortices due to each CS term and show that the new CS term dictates a universal Abelian exchange statistics phase of the vortices equal to $e^{i\pi/8}$. In contrast, for neutral superfluids, the partial CS term spoils the quantization of the exchange phase by adding a long-distance non-universal correction. For charged superfluids, screening effects exponentially diminish the latter over the effective penetration depth. This sets a low bound for the distance between vortices during exchange processes required for topological quantum computation.

\section{Gauge transformation}
We start with the action for a spinless chiral $p$-wave superconductor~\cite{kleinert1978collective},
%%%%%
$
Z=\int\mathcal{D}(\bar{\eta},\eta)e^{i S},
$
%%%%%
where $\eta=\left(\phi,\bar{\phi}\right)^\intercal$ and $\bar{\eta}=\left(\bar{\phi},\phi\right)$ are the Nambu spinors with Grassmann variables $\phi(r)$ and $\bar\phi(r)$ in the coordinate space $r=(\mathbf{r},t)$. In the following, we will interchangeably use $z\equiv t$ as the third coordinate and $d^3 r = d{\bf r}d t$. The action is
%%%%%
$
S=\frac{1}{2} \int d^3r\; \bar{\eta}\mathcal{G}^{-1}\eta,
$
%%%%%
with $\mathcal{G}^{-1}=i\partial_t-\mathcal{H}$ the inverse Green's function matrix and the Bogoliubov--de Gennes Hamiltonian density~\cite{ivanov2002random},
%%%%%
\begin{equation}\label{eq:hamiltonian}
%\mathcal{H}=  \tau_3\left( \frac{(\mathbf{p}-\tau_3\mathbf{A})^2}{2 m}-\epsilon_F -A_0 \right)-\tau_1\lbrace\Delta,p_1\rbrace -\tau_2\lbrace\Delta,p_2\rbrace
%\mathcal{H}=  \tau_3\left( \xi_{\mathbf{p}-\tau_3\mathbf{A}}-A_0 \right)-\tau_1\lbrace\Delta,p_1\rbrace -\tau_2\lbrace\Delta,p_2\rbrace.
\mathcal{H}= \left(\begin{array}{cc}
\xi_{\mathbf{p}-\mathbf{A}}-A_t & e^{i\theta/2}\Delta(\mathbf{p})e^{i\theta/2} \\
e^{-i\theta/2}\Delta(\mathbf{p})^\dagger e^{-i\theta/2} & -\xi_{\mathbf{p+\mathbf{A}}}+A_t
\end{array}\right).
\end{equation}
%%%%%
Here, $\xi_{\mathbf{p}}$ is the dispersion of excitations above the ground state, $\mathbf{p} = -i\boldsymbol{\nabla}$ is the momentum operator, $\Delta(\mathbf{p})$ is the amplitude and $e^{i\theta(\mathbf{r},t)}$ is the phase of the superconducting order parameter (including vortices), and $A=(\mathbf{A},A_t)$ is the  electromagnetic gauge field. (In a neutral superfluid, $A=0$.)
We assume $e=c=\hbar=1$. In the continuum, $\xi_{\mathbf{p}} = \mathbf{p}^2/2m-\epsilon_F$ with $\epsilon_F$ the Fermi energy and $\Delta(\mathbf{p})=v(p_x+ip_y)$ with $v$ the slope of the pairing order parameter in momentum space.  

In order to keep track of the winding number around each vortex we define $\theta(\mathbf{r},t)=\sum_{j=1}^n \theta_j(\mathbf{r},t)$, 
where $\theta_j=\arg(\mathbf{r}-\mathbf{x}_j)\in(2\pi\ell_j,2\pi(\ell_j+1)]$ is the phase around the vortex located at $\mathbf{x}_j(t)$ and $\ell_j$ is its winding number. We take the branch cut of $\arg(\mathbf{r})$ to be the positive real axis and index the corresponding Riemann sheets with the branch number $\ell\in\mathbb{Z}$~\cite{Note1}.

The partition function is invariant under a unitary transformation, $U$, of the inverse Green's function with a Jacobian of unit modulus; that is, $U=e^{i\alpha_0}e^{i\alpha^\mu\tau_\mu}$, where $\tau_x$, $\tau_y$ and $\tau_z$ are Pauli matrices in the Nambu space. We demand that $U$ respect the particle-hole symmetric structure of the spinor fields. This means that $U$ must transform $(\bar\eta,\eta)$ in such a way that ensures one spinor remains the conjugate transpose of the other and each component of the spinor is the conjugate of the other. The requirement is equivalent to the condition $U^\dagger=\tau_x U^\intercal\tau_x$. In the operator language, this is the condition to maintain the fermionic commutations relations under the Bogoliubov transformation. One can readily show that any such $U$ is composed of a finite product of the following matrices: $\tau_x$, $\tau_y$, $e^{i\mu\tau_z}$ and $e^{i\pi m}\mathbb{1}$, where $\mu \in \mathbb{R}$ and $m \in \mathbb{Z}$. The actual number of distinct sequences can be reduced through use of the commutations relations between the generators and is ultimately finite.

To proceed further, it is convenient to gauge away the phase of the superconducting order parameter.
This will add space-time gradients of $\theta(\mathbf{r},t)$ to the electromagnetic potential in the kinetic term.
A naive transformation, $e^{i\tau_z\theta/2}$, which involves only the phase of the order parameter, leads to multi-valuedness in the presence of vortices. 
To avoid this problem Anderson~\cite{anderson1998anomalous} suggested using the transformations $e^{-i(\tau_z \mp \mathbb{1}) \theta/2}$, resulting in the superfluid velocity appearing as an effective gauge field in either the electron or the hole component of the Hamiltonian. This gauge choice becomes possible when opposite spins are associated with the two components of the Nambu spinor. Franz and Te\v{s}anovi\'c~\cite{franz2000quasi,vafek2001quasi} developed the transformation $e^{i(\tau_z + \mathbb{1}) \theta_\text{A}}\cdot e^{i(\tau_z - \mathbb{1}) \theta_\text{B}}$ for a periodic bipartite vortex lattice, where A and B are the two sublattices. The vortices should be assigned to the subsets in such a way that the effective magnetic field vanishes on average. Physically, a vortex assigned to subset A will be seen by electrons and be invisible to holes, while vortex assigned to subset B will be seen by holes and be invisible to electrons. Inevitably, in both transformations, particle-hole symmetric structure of the spinors cannot be maintained without additional constraints on the ensemble of allowed partitions of $\theta$.

Instead, we suggest the following transformation:
%%%%%
\begin{equation}\label{eq:U}
U=e^{i \tau_z \theta(\mathbf{r},t)/2}e^{i\gamma(\mathbf{r},t)},
\end{equation}
%%%%%
where $\theta$ is the phase function and $\gamma=\pi\sum_j\ell_j$ keeps the transformation properly single-valued by supplying the required sign each time the winding number in $\theta$ changes as it evolves in space and time. Our transformation is similar in spirit to the Franz-Te\v{s}anovi\'c transformation, especially as formulated in Ref.~\onlinecite{sheehy2003feynman}, but it manifestly preserves the particle-hole symmetry of the action.
Upon applying this gauge transformation, two gauge fields appear in the action: the $a_\nu = A_\nu-\partial_\nu\theta/2$ couples only to the kinetic energy terms, with opposite signs for particles and holes; and the $b_\nu = \partial_\nu\gamma$ couples minimally to momentum, both in the kinetic energy and in the pairing term. We note that the $b$ gauge field is associated with the vortex branch cuts and its corresponding current is proportional to the vortex current.  After this transformation, we find
%%%%%
\begin{equation}
\mathcal{G}^{-1} = i\partial_t - b_t + \tau_z a_t - \tau_\mu h_\mu(\mathbf{p}-\mathbf{b},\mathbf{a}),
\end{equation}
%%%%%
where the 3-vector
$h(\mathbf{p},\mathbf{a}) = (\Re\Delta(\mathbf{p}), \Im\Delta(\mathbf{p}),\xi_{\mathbf{p}-\tau_z\mathbf{a}})$.\\

\section{Effective action}
We can now integrate out the fermion fields to find the effective action, $S_{\text{eff}}=\frac{i}{2}\text{Tr} \ln\mathcal{G}$, where $\text{Tr}(\cdot)$ stands for $\int d\mathbf{r}d t \langle\mathbf{r},t\left| \text{tr}(\cdot)\right|\mathbf{r},t\rangle $ and ``$\text{tr}$'' is the trace over the Nambu space.
A tedious but straightforward calculation yields (see Appendix), to second order in the gauge fields,
%%%%%
\begin{multline}\label{eq:action}
S_{\text{eff}}= \int d\mathbf{r}d t \left[ na_t + \rho_t a_t^2 - \rho_{ij} a_ia_j 
\right.
\\ \left. - \frac{\kappa_a}{8\pi} \varepsilon_{tij} a_t\partial_i a_j  + \frac{\kappa_b}{8\pi}\varepsilon_{\lambda\mu\nu} b_\lambda \partial_\mu b_\nu\right],
\end{multline}
%%%%%
where $\varepsilon_{\lambda\mu\nu}$ is the antisymmetric tensor and latin indices $i,j$ run over the spatial components.
The coefficients appearing in Eq.~(\ref{eq:action}) are found in terms of $g(\mathbf{k})\equiv h(\mathbf{k},0)$ as follows: 
%%%%%
\begin{eqnarray}
n &=& \frac1{8\pi^2}\int d\mathbf{k}\left(1-\frac{g_z}{|g|}\right); \\
\rho_t &=& \frac1{16\pi^2}\int d\mathbf{k}\frac{g_x^2+g_y^2}{|g|^3}; \\
\rho_{ij} &=& \frac1{16\pi^2}\int d\mathbf{k}\left(1-\frac{g_z}{|g|}\right)\partial_{k_i}\partial_{k_j}g_z.
\end{eqnarray}
%%%%%
Note that $n$ is just the superfluid density. The coefficient of the partial CS term for $a$,
%%%%%
\begin{equation}
\kappa_a = \frac1{4\pi}\int \frac{\varepsilon_{i\nu\lambda}g_i\partial_{k_x}g_\nu\partial_{k_y}g_\lambda}{|g|^3} d\mathbf{k},
\end{equation}
%%%%%
is non-universal and depends on the details of the system. The coefficient of the full CS term for $b$, on the other hand,
%%%%%
\begin{equation}
\kappa_b = \frac1{4\pi}\int \frac{\varepsilon_{\mu\nu\lambda}g_\mu\partial_{k_x}g_\nu\partial_{k_y}g_\lambda}{|g|^3} d\mathbf{k},
\end{equation}
%%%%%
is the Pontryagin charge of the field $g_\mu(\mathbf{k})$ and is therefore always an integer. The action in Eq.~(\ref{eq:action}) is our central result. 

In the continuum limit, we have $\xi_{\mathbf{k}}=\mathbf{k}^2/2m-\epsilon_F$ and $\Delta(\mathbf{k})=v(k_x+ik_y)$. Calculating the coefficients in this limit, we find the following values:
$n\sim (m v)^2\ln \left(\frac{\Lambda }{mv ^2}\right)$ with  $\Lambda$ an energy cut-off; $\rho_t = m\kappa_a^\infty/4\pi$; and $\rho_{ij}=(n/2m)\delta_{ij}$, which reflects the Galilean invariance in the continuum~\cite{sonin2013transverse}.
The CS coefficients in the continuum limit are
%%%%%
\begin{eqnarray}
\kappa_a^\infty &=& \left[1-2 \frac{\epsilon_F}{mv^2}\Theta(-\epsilon_F)\right]^{-1}, \\
\kappa_b^\infty &=& \Theta(\epsilon_F),
\end{eqnarray}
%%%%%
where $\Theta$ is the step function.  Note that this extends the results obtained in Refs.~\cite{goryo2000vortex,stone2004superfluidity} to the strong pairing regime, $\epsilon_F<0$. 

\begin{figure}
\includegraphics[width=3.4in]{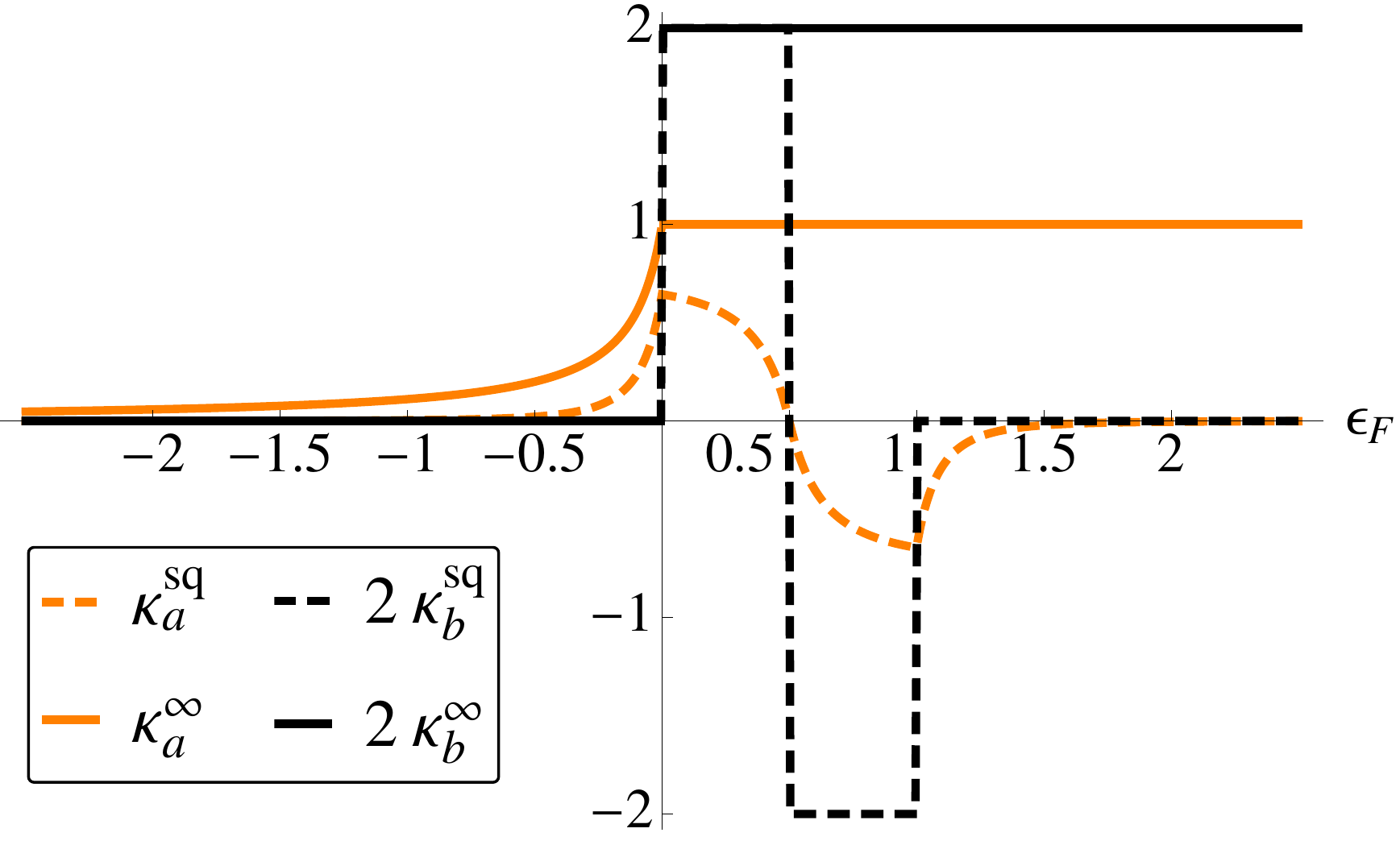}
\caption{{ Coefficients of the Chern-Simons terms}. The coefficients $\kappa_a$ and $\kappa_b$ of the partial (orange) and full (black) Chern-Simons terms are shown for a system in the continuum (solid) and on the square lattice (dashed) for $v=m$. For clarity we show $2\kappa_b$. The Fermi energy is in units of $4/(md^2)$, the bandwidth of the square lattice, with lattice spacing $d$. The exchange angles due to each term can be obtained by multiplying the plotted values with $\mp\pi/16$.}
\label{fig1}
\end{figure}

For comparison, we have also calculated these coefficients for a system on the square lattice. In this case, 
$\xi_{\mathbf{k}} = \frac1{m d^2}(2-\cos k_x d-\cos k_y d)-\epsilon_F$ and $\Delta(\mathbf{k})=\frac vd(\sin k_xd+i\sin k_y d)$, where $d$ is the lattice spacing.
The coefficients $\kappa_{a,b}^{\infty,\text{sq}}$ are plotted in Fig.~\ref{fig1} as a function of $md^2\epsilon_F/4$. We observe that $\kappa_b^{\text{sq}}$ acquires the values $\pm 1$ in the topological regime $0<\epsilon_F<4/(md^2)$~\cite{bernevig2013topological} and zero otherwise. In contrast, $\kappa_a^{\infty}$ and $\kappa_a^{\text{sq}}$ are clearly non-universal and vary with $\epsilon_F$, showing derivative discontinuities when crossing into the topological regime. The sign change of $\kappa_a^{\text{sq}}$ on the lattice signals a sign reversal in the Hall response of the superconductor~\cite{stone2004superfluidity,huber2011topological}.

%-------- Fig 1 --------%

\section{Vortex dynamics and exchange}
The effective action, Eq.~(\ref{eq:action}), now captures correctly the physics of vortices. This is exemplified by the physical significance of each term appearing in the action. The first term gives rise to the Magnus force on a moving vortex. To see this, note that for a moving vortex $\partial_t\theta= -\dot{\mathbf{x}}\cdot\boldsymbol{\nabla}\theta$, where $\mathbf{x}(t)$ is the position of the vortex. So, the first term yields $-\int dt\; \dot{\mathbf{x}}\cdot \boldsymbol{\mathcal{A}}_M$ with $\boldsymbol{\mathcal{A}}_M=-\int d\mathbf{r}\;n\boldsymbol{\nabla}\theta/2$. Therefore, the vortex is subject to a Lorentz-like force $\dot{\mathbf{x}}\times\boldsymbol{\mathcal{B}}_M$ where the Magnus flux $\boldsymbol{\mathcal{B}}_M=\boldsymbol{\nabla}_{\mathbf{x}}\times\boldsymbol{\mathcal{A}}_M=\pi n\hat{\mathbf{z}}$ is proportional to the superfluid density. The contribution from the electromagnetic gauge field $A_t$ in a superconductor vanishes due to the overall charge neutrality of the system~\cite{ao1993berrys}.
The second and third terms, in conjunction with the Maxwell Lagrangian, give rise to the usual screening of vortices through the Meissner effect. The second term also contributes to the mass of the vortex by generating a term $\int d t \frac12 m_v \dot{\mathbf{x}}^2$ in the action, where $m_v=\int d\mathbf{r}\rho_t(\boldsymbol{\nabla}\theta/2-\mathbf{A})^2$.

The fourth and fifth terms, as we now show, carry significant information about the dynamics of vortices. Previous work on the effective low-energy theory of the $p$-wave superconductor, using only the $\mathbb{U}(1)$ part of our transformation~\cite{volovik1987analog,goryo2000vortex}, yielded an action similar to that of an s-wave superconductor but with an additional partial CS term.  Stone and Roy~\cite{stone2004superfluidity} attributed this partial CS term to the existence of a Hall-like response to external fields. They recognized that the Hall current depends on the external field primarily through its effect in modifying the density. 
Note that the partial CS term we derive here is different from the one appearing in the literature, since in our case $\boldsymbol{\nabla}\times\boldsymbol{\nabla}\theta$  is explicitly nonzero due to the presence of vorticity in $\theta$. Moreover, the full CS term derived here is entirely absent in previous work. As we show now, both of these terms have significant contributions to the exchange statistics of vortices.

Vortices in chiral $p$-wave superconductors are known to obey non-Abelian statistics, the mechanism behind which relates to the Majorana zero modes localized at their cores. In the presence of $2n$ vortices, the ground state of the system is $2^{n}$-fold degenerate. This degenerate ground state is further divided into two sectors of definite parity $(-1)^{N}=\pm1$, where $N$ is the total fermion number. The full braid statistics of vortices can be written using three matrices: $R$; $F$; and $B$. Roughly speaking, $R$ specifies the exchange of two vortices when their fusion outcome is known, $F$ specifies the associativity of the exchange among three anyons, and $B=F^{-1}RF$ is the generator of the full braid group
of the vortices in the model. The possible choices of $R$ and $F$
are constrained by consistency relationships. These matrices have been computed for a chiral $p$-wave superconductor by Ivanov \cite{ivanov2001non-abelian} and found to be, up to an overall phase, proportional to those in the Ising anyon model. In this model,
a vortex, $\sigma$, and antivortex, $\overline{\sigma}=\sigma$,
fuse according to the fusion rule $\sigma\times\sigma=I+\psi$, where
the fusion channels $I$ and $\psi$ are, respectively, the vacuum
(boson) and fermion. In this basis, the $F$-matrix is real and is
given by $F=\frac{1}{\sqrt{2}}\left(\sigma_z+\sigma_x\right)$, while $R=e^{-i\chi}\,\text{diag}(1,i)$. The phase $\chi=\pi/8$ is fixed in this model by consistency relations between $R$ and $F$~\cite{pachos2012introduction}. 
%In the field theoretical language, the exchange statistics of $N$ Majorana fermions is related to their dynamical term and within the Born-Oppenheimer approximation gives rise to an SO($N$) gauge structure operating in the space of degenerate ground states. 
However, without a full calculation of $\chi$ in the chiral $p$-wave superconductor, one cannot make a meaningful connection to the Ising anyon model. 

\begin{figure}[t]
\includegraphics[width=3.4in]{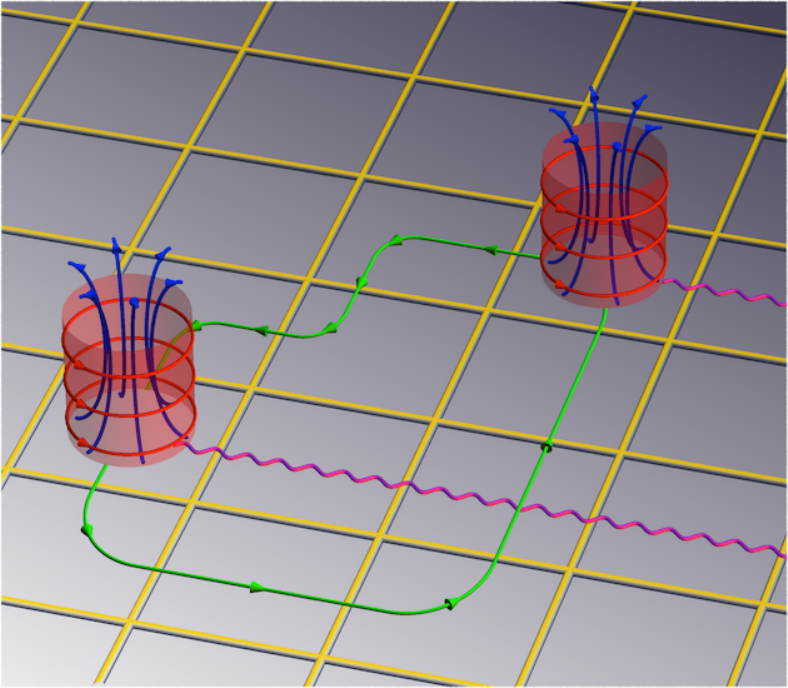}
\caption{{ The exchange scheme of two vortices}. The exchange path is shown in green and the branch cuts in purple. In a charged superfluid, the magnetic field, blue, is screened by the supercurrent, red. The calculation is simplified when the path is a semicircle of one vortex around the other followed by two radial displacements.}\label{fig:exch}
\label{fig2}
\end{figure}

Our strategy in this work is to calculate $\chi$ by performing a monodromy, which describes a full encircling of one vortex around the other. A general argument shows that the monodromy in the vacuum fusion channel is $R^{2}=e^{-2 i \chi}$~\cite{pachos2012introduction}. This calculation may be done in the same ground state without complications due to the ground-state degeneracy. In our field theory, the monodromy is the Berry's phase in the matrix element of the evolution operator for the exchange of two vortices in the even-parity ground state~\cite{wilczek1990fractional}. See Fig.~\ref{fig:exch} for illustration.

%Abelian exchange statistics characterizes the manner in which the wavefunction transforms under the interchange of indistinguishable excitations. Specifically, the statistics is reflected in the wavefunction as a boundary condition, $\left|\psi(\mathbf{x}_1,\mathbf{x}_2)\right\rangle = e^{i\chi}\left|\psi(\mathbf{x}_2,\mathbf{x}_1)\right\rangle$,with $\mathbf{x}_1$ and $\mathbf{x}_2$ the positions of the excitations and $e^{i\chi}$ the phase accumulated during their exchange. The exchange angle $\chi$ can be obtained by evaluating the Berry's phase accumulated upon the exchange of two particles~\cite{wilczek1990fractional}. A full CS term in the Lagrangian produces a quantized Berry's phase that depends solely on the topology of the ground state and has long been used to describe fractional exchange statistics of excitations. 

%%-------- Fig 2 --------%

At first sight, the $\mathbb{Z}_2$ nature of the $b$ gauge field in our effective theory seems to make the calculation of the Berry's phase due to the full CS term tricky. However, this situation is similar to the situation encountered in the singular string gauge of the more common $\mathbb{U}(1)$ gauge theory, in which the gauge field is zero everywhere except on a string emanating from the vortex. One may show that the string gauge is continuously connected to a smooth gauge without changing the winding numbers along the process. Therefore, we can calculate the Berry's phase contribution of the $b$ gauge field in the usual way by writing $b=b_1+b_2$, where $b_1$ and $b_2$ are associated with the two vortices, and considering the cross terms between them. Both cross terms contribute equally since, by partial integration, $\int \varepsilon_{\lambda\mu\nu}b_{1\lambda}\partial_\mu b_{2\nu}=\int \varepsilon_{\lambda\mu\nu}b_{2\lambda}\partial_\mu b_{1\nu}$. Assuming for simplicity that only vortex 2 is moving, we have $\varepsilon_{\lambda\mu\nu}\partial_\mu b_{1\nu}=\pi\delta(\mathbf{r})\delta^\lambda_t$, and
%%%%%
\begin{equation}
\label{eq:berry2}
\chi_b = \frac{\kappa_b}{4}\int d\mathbf{r}dt\: \delta(\mathbf{r})b_{2t} =\frac{\pi \kappa_b}{8},
\end{equation}
%%%%%
which is, as advertised, quantized in the weak pairing (topological) regime to the value $\pi/8$.

The partial CS term in~(\ref{eq:action}) also contributes to the Berry's phase, albeit not in a quantized fashion due to the non-universal behavior of $\kappa_a$. 
We write again $a=a_1+a_2$ for two vortices and consider the cross terms in the CS term between $a_1$ and $a_2$. In a superfluid the external electromagnetic gauge field is absent and we have $a_{1\mu}=-\frac12\partial_\mu\arg(\mathbf{r}-\mathbf{x}_1(t))$,  where $\mathbf{x}_1(t)$ is the position of vortex 1, and similarly for $a_2$. The calculation is simplified by assuming that only vortex $2$ moves, so that $a_{1t}=0$. Then, only one of the cross terms contributes and
%%%%%
\begin{equation}
\begin{aligned}
	\label{eq:berry}
	\chi_a^{\text{sf}}=-\frac{\kappa_a}{8\pi}\int d\mathbf{r} dt\;a_{2t}(\boldsymbol{\nabla}\times \mathbf{a}_1)_t=-\frac{\pi \kappa_a}{16}.
\end{aligned}
\end{equation}
%%%%%
In a neutral superfluid, this leads to a non-universal long range contribution to the exchange phase of vortices. 

By contrast, for a charged superfluid, the screened magnetic field is screened as $(\boldsymbol{\nabla}\times\mathbf{A})_t=\hat{\mathbf{z}}\:K_0(r/\lambda)/(2\lambda^2)$, associated with a vortex at the origin, where $K$ is the modified Bessel function of the second kind and $\lambda$ is the (effective) penetration depth. This modifies the result by a geometric phase,
%%%%%
\begin{equation}
\frac{\kappa_a}{8\pi}\int d\mathbf{r} dt\;a_{2t}(\boldsymbol{\nabla}\times\mathbf{A}_1)_t=
-\frac{\pi \kappa_a}{16}\left[1- \frac{R_0}{\lambda} K_1\left(\frac{R_0}{\lambda}\right)\right],
\end{equation}
%%%%%
for a circular exchange at distance $R_0$. So,  in a superconductor the total exchange angle due the partial CS term is
%%%%%
\begin{equation}
\chi_a^{\text{sc}}=-\frac{\pi\kappa_a R_0}{16 \lambda}K_1\left(\frac{R_0}{\lambda}\right).
\end{equation}
%%%%%
When the distance between the vortices is much longer than $\lambda$ this exchange angle vanishes exactly. However, at distances smaller or comparable to $\lambda$, non-universal contributions to the exchange phase will occur.

Therefore, the total exchange angle $\chi=\chi_a+\chi_b$ depends on the details of the dispersion and, in particular, is different in a chiral $p$-wave superfluid from that in a superconductor due to screening effects.\\

\section{Discussion}
In the derivation above we concentrated on the case where the size of the core of the vortex is vanishingly smaller compared to other length scales. We note that even in this limit, the $2\pi$ winding of the phase entails the presence of the protected zero mode in the topological phase. In addition, higher-energy subgap states may occur, localized at the vortex core~\cite{seradjeh2008midgap}. The field theory presented above would include the effects of both the zero mode and the subgap states in the two gauge fields $a$ and $b$ if all orders of the loop expansion are retained. To the second order, we find only the Chern-Simons term, which fully encodes the topological exchange phase. This phase is quantized and cannot be modified without closure of the gap. Non-universal effects associated with the subgap states may occur in higher order in perturbation theory, which may include effects such as population transfer between closely separated intra-vortex states~\cite{moller2011structure}. 

Taking a finite core size may allow additional localized sub-gap states to get trapped within the vortex. One can model this case by varying the chemical potential $\epsilon_F$ around the vortex through the topological phase transition between the topological weak-pairing phase ($\epsilon_F>0$) outside of the vortex core and the non-topological strong-pairing phase ($\epsilon_F<0$) within the vortex core. As far as topological properties are concerned, this is equivalent to taking the order parameter to zero at the vortex core but lends itself better to field theoretical analysis~\cite{read2000paired,nayak2008non}. In this description, the loci of $\epsilon_F=0$ that encircle the vortices' cores are internal edges of the system and accommodate gapless excitations. Although their proximity to the Majorana zero modes may affect the coherence of the vortices~\cite{elliott2014colloquium,moller2011structure,ivanov1999the}, as long as they do not mix with the continuum states the Majorana zero mode remains intact~\cite{akhmerov2010topological}. To incorporate this into our field theory, we take the chemical potential to be $\mu(r)=\epsilon_F+\delta\mu(r)$, where  $\delta\mu(r)$ denotes the deviation from $\epsilon_F$ and has support mostly within the vortex core, i.e., within the coherence length. The new term can be absorbed into $a_t\to a_t - \delta\mu$. One new term that appears in the field theory, $\delta\mu(r)n$, pushes vortices to diffuse along the chemical potential gradient occurring due to other vortices when their cores overlap. A second term $\propto\delta\mu(r) \epsilon_{ij}\partial_i a_j$ generates energetic contributions which go to zero at distances that are larger than the coherence or penetration lengths (whichever is larger). Since there is no coupling between $a$ and $b$ in the effective theory, this modification does not change the topological CS term. At distances larger than the coherence length and to second order in perturbation theory, we find no contribution to the topological Abelian phase associated with the exchange of vortices.

%\subsection{Consequences for the non-Abelian anyon model for vortices}
%***. Therefore, our results in the previous section show that for the p-wave superfluid, $\theta=\pi/8+\theta_{\text{non-U}}$, where the non-universal contribution, $\theta_{\text{non-U}}$, is a deviation from the pristine Ising anyon description. For a charged superfluid this contribution is screened and distant vortices exponentially approach the Ising universality class.
It is also illuminating to compare our results to the non-Abelian Moore-Read state, which is one of the prominent candidate wavefunctions describing the quantum Hall plateau at filling factor $5/2$ \cite{moore1991nonabelions,read2000paired}. While lying in the same universality class of Ising anyons as chiral $p$-wave superfluids, the Moore-Read state is realized at large magnetic fields, leading to the appearance of an additional Chern-Simons term in the mean-field action. This extra term endows the quasi-particles with an $e/4$ charge and half a flux quantum of fictitious magnetic field. Consequently, there is an additional $\pi/8$ exchange phase for quasi-particles which should be added to the pure Ising $\pi/8$ contribution, for a total exchange phase of $\pi/4$. Non-universal deviations from this value must be governed by the magnetic length, so any mapping of our results to the Moore-Read state, if it at all exists, remains to be worked out. In contrast to the Moore-Read state, our main result here demonstrates that vortices in a screened chiral $p$-wave superconductor could realize a pristine Ising anyon model.

\section{Conclusion}

We have derived an effective action of vortices in a spinless chiral $p$-wave superfluid by properly treating the vortex branch cuts and revealed an Abelian $\mathbb{Z}_2$ gauge structure for the chiral $p$-wave superfluid. In principle, our transformation is applicable to any pairing symmetry and arbitrary distribution of vortices. In the $s$-wave case, we have checked that this does not produce additional terms in the action. In the $d$-wave case a similar approach has been used to formulate an effective theory of cuprate superconductors~\cite{franz2001algebraic,herbut2002qed3}, but no CS term was found. 

The topological quantum computation scheme relies on adiabatic braiding of non-Abelian anyons to generate the quantum computation. Among non-Abelian anyon models. Majorana fermions are arguably the closest to experimental work. However, braiding of vortices carrying Majorana fermions is non-universal unless supplemented by a missing $\pi/8$ gate. While this gate can be generated by sacrificing topological protection it remains of fundamental importance to provide a proof-of-principle topological scheme to supply the missing $\pi/8$ gate, thus avoiding costly error protection protocols. The results presented here allow the realization of the missing $\pi/8$ gate through multiple braiding of the anyons~\cite{bonderson2013twisted}. As argued above, such braidings should be performed at distances larger than both the coherence length and the screening length.

In this work, we restricted our attention to the Abelian gauge transformations~(\ref{eq:U}). This is enough to infer the Abelian exchange phase of vortices. It can also be used to deduce the existence of zero energy Majorana modes. Using the particle-hole symmetry of the Hamiltonian density, we can write the number density of zero modes as $\nu_0=2\left< \bar\eta(r)\eta(r) \right>$~\cite{herbut2007zero,lu2014zero}. Now, since
$$
\left< \bar\eta(r)\eta(r) \right> = 2\left<\delta S_{\text{eff}}/\delta b_t\right>=\frac{\kappa_b}{2\pi}(\boldsymbol{\nabla}\times\mathbf{b})_t,
$$
and $b$ is defined as a $\mathbb{Z}_2$ gauge field, we find
\begin{equation}
\nu_0 = \kappa_b\sum_j\delta(\mathbf{r}-\mathbf{x}_j(t)) \quad (\mathrm{mod}~2),
\end{equation}
which is quantized and equal to the single winding vortex density (mod 2) in the weak pairing regime. A natural question for future work is whether the other parts of the full group of gauge transformations harbor additional physics. Indeed, as is well known, the zero energy Majorana modes endow the vortices with the non-Abelian statistics of Ising anyons~\cite{read2000paired,ivanov2001non-abelian}. It would be interesting to see if such a non-Abelian representation emerges in the gauge structure of the effective vortex action by using the entire group of gauge transformations.

\section*{Acknowledgments}
This work has been supported in part by the Israel Science Foundation (Grant No. 401/12), the European Union's Seventh Framework Programme (FP7/2007-2013) under Grant No. 303742, and the College of Arts and Sciences, Indiana University. This material is based upon work supported by the National Science Foundation under Grant No. DMR-1350663. Initial support by the NSERC of Canada (BS) and ICMT at University of Illinois, Urbana-Champaign is also acknowledged.

\appendix*
\section{Derivation of the action}

In the following, we provide the details of the derivation of the effective action appearing in the Results section, Eq.~(\ref{eq:action}). We first write the `dressed' Green's function as $\mathcal{G}^{-1}=\mathcal{G}_0^{-1}+V$, where the `bare' Green's function is ${\mathcal{G}_0(k)=(k_t\mathbb{1}-g({\mathbf k})\cdot\tau)^{-1}}$ and $V$ depends explicitly on $a$ and $b$, i.e. 
\begin{equation}
{V=-b_t+\tau_z a_t-\tau_\mu h_\mu({\mathbf p}-{\mathbf b},{\mathbf a})+\tau_\mu h_\mu({\mathbf p},0)}.
\end{equation}
We now perform a perturbative expansion to second order in the gauge fields, writing 
\begin{eqnarray}\nonumber
	S_{\text{eff}} &=& -i \ln \text{Pf}(\mathcal{G}^{-1}) = -i\frac{1}{2}\text{Tr}\ln(\mathcal{G}^{-1})\\
	&=& -\frac{i}{2}\text{Tr} \ln\left(\mathcal{G}_0^{-1}\right)-\frac{i}{2}\text{Tr} \ln(1+\mathcal{G}_0 V)\\
	&\simeq& -\frac{i}{2}\text{Tr} \ln\left(\mathcal{G}_0^{-1}\right)-\frac{i}{2}\text{Tr} (\mathcal{G}_0 V)+\frac{i}{4}\text{Tr}(\mathcal{G}_0 V\mathcal{G}_0 V).\nonumber
\end{eqnarray}
The fields $a$ and $b$ couple via their associated currents
\begin{eqnarray}
	\nonumber j_a^\mu &=& \delta^\mu_{t}\tau_z+\partial_{k_\mu} g_z(1-\delta^\mu_t),\\
	j_b^\mu &=& \partial_{k_\mu}\mathcal{G}_0^{-1}.
\end{eqnarray}
To calculate traces, we use the following formulas
\begin{eqnarray}
	\nonumber&\text{tr}\{\tau_\mu \tau_\nu \}=2 \delta_{\mu\nu},\\
	&\text{tr}\{\tau_\lambda\tau_\mu\tau_\nu\}=2i\epsilon_{\lambda\mu\nu},\\
	\nonumber&\text{tr}\{\tau_\lambda\tau_\mu\tau_\nu\tau_\sigma\}=2(\delta_{\lambda\mu}\delta_{\nu\sigma}-\delta_{\lambda\nu}\delta_{\mu\sigma}+\delta_{\lambda\sigma}\delta_{\mu\nu}).
\end{eqnarray}

\subsection{The non-vanishing terms}

We proceed to derive the coefficients of the five terms appearing in the action, Eq.~(\ref{eq:action}).\\

\begin{figure}[t]
\includegraphics[width=3.4in]{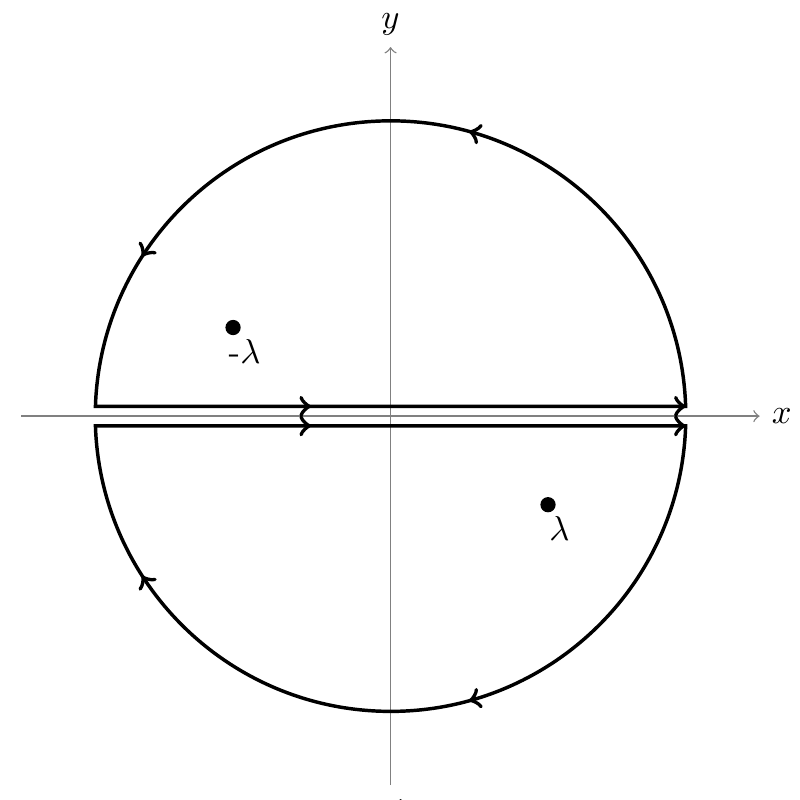}
\caption{{ Contours of integration and poles of the Green's function}. The contours of integration that are used in zero temperature calculations for the particle (top) and hole (bottom) part of the Green's function.}
\label{fig:S1}
\end{figure}

{\bf The coefficient $\boldsymbol{n}$}. %%%%%%% n
The coefficient multiplying $a_t$ is ${n=-\frac{i}{2(2\pi)^3}\int d^3 k\, \mathrm{tr}\left(\mathcal{G}_0\tau_z\right)}$. Since it contains an integration over a single Green's function, care should be taken in its calculation. The correct analytical structure requires that the Green's function is multiplied by an exponent $e^{i\tau_z k_t \eta}$, where $\eta\to 0$, leading to the expression
\begin{equation}
n=-\frac{i}{2(2\pi)^3} \sum_{s=\pm}\int{d^3k}\frac{s k_t+g_z}{k_t^2-|g|^2+i\eta}e^{i s \eta k_t}.
\end{equation} 
Using contour integration over $k_t$ (see Fig.~\ref{fig:S1} where $\lambda\equiv\sqrt{|g|^2-i\eta}$) one obtains the expression 
\begin{eqnarray}
	n=\frac{1}{8\pi^2}\int d{\mathbf{k}}\left(1-\frac{g_z}{|g|}\right).
\end{eqnarray}
In the $p$-wave case the integral is formally divergent and an energy cutoff $\Lambda=\Lambda_k^2/(2m)-\epsilon_F$ needs to be introduced (here $\Lambda_k$ is a  momentum cutoff set only by the inverse lattice spacing)
\begin{equation}
n(\Lambda)=\frac{m}{4\pi}\int\limits_{-\epsilon_F}^{+\Lambda}d\xi\left(1-\frac{\xi}{\sqrt{\xi^2+2mv^2(\xi+\epsilon_F)}}\right).
\end{equation}

%-------- Fig 3 --------%

{\bf The coefficient $\boldsymbol{\rho_t}$}. %%%%%%% \rho_t
Writing the appropriate second order correlator,
\begin{multline}
\rho_t =\frac{i}{32\pi^3}\int{d^3k} \text{ tr}\left(\mathcal{G}_0 j^t_a \mathcal{G}_0 j^t_a \right)\nonumber\\ 
=\frac{i}{16\pi^3}\int{d^3k}\frac{k_t^2-g_x^2-g_y^2+g_z^2}{(k_t^2-|g|^2+i \eta)^2}\nonumber\\
=\frac{1}
{16\pi^2}\int{ d\mathbf{k}}\frac{g_x^2+g_y^2}{|g|^3},
\end{multline}
where we used the integral
\begin{equation}
	\label{eq:integral}
	\int_{-\infty}^{\infty}dk_t\frac{\alpha k_t^2+\beta}{(k_t^2-|g|^2+i\eta)^2}=\frac{i\pi (-\alpha |g|^2+\beta)}{2|g|^3},
\end{equation}
with $\alpha=1$ and $\beta=g_z^2-g_x^2-g_y^2$. For $p$-wave superfluids in the infinite system limit,
\begin{eqnarray}
	\rho_t=\frac{1}{16\pi^2}\int d\mathbf{k}\frac{v^2 {\mathbf k}^2}{(\xi_{\mathbf{k}}^2+v^2 \mathbf{k}^2)^{3/2}}=\frac{m \kappa_a^{\infty}}{4\pi},
\end{eqnarray}
where $\kappa_a^{\infty}=\left(1-\frac{\epsilon_F-|\epsilon_F|}{m v^2}\right)^{-1}$ coincides with the coefficient of the partial CS term, to be derived below. \\ 

{\bf The coefficient $\boldsymbol{\rho_{ij}}$}. %%%%%%% \rho_ij
Formally, this coefficient has contributions both from first order and second order in the gradient expansion. The first-order contribution is
\begin{equation}
\frac{-i}{2(2\pi)^3}\int{d^3k}\text{ tr}\left[\mathcal{G}_0 \left(-\frac{\tau_z}{2m}\delta_{ij}\right)\right]=-\frac{n}{2m}\delta_{ij},
\end{equation}
For $g_z=\xi$, we can write $\delta_{ij}/m=\partial_{k_i}\partial_{k_j}g_z$, to obtain the form in the main text. The second order contribution exactly vanishes following the integration over $k_t$,
\begin{multline}
\frac{i}{32\pi^3}\int{d^3k} \text{ tr}\left(\mathcal{G}_0 j_a^i \mathcal{G}_0 j_a^j \right)=\\ \frac{i}{16\pi^3}\int{d^3k}\frac{\partial g_z}{\partial{k_i}}\frac{\partial g_z}{\partial{k_j}}\frac{k_t^2+|g|^2}{(k_t^2-|g|^2+i\eta)^2}=0.
\end{multline}

{\bf The coefficient $\boldsymbol{\kappa_a}$}. %%%%%%% \kappa_a
To calculate $\kappa_a$ we consider the correlator of $j_a^t$ and $j_a^j$ to first order in $q_i$ (no summation convention)
\begin{align}
&{}\frac{iq_i}{64\pi^3}\int{ d^3k}\, \text{tr}\left[\partial_{k_i}\mathcal{G}_0 \tau_3 \mathcal{G}_0 \frac{\partial g_z}{\partial k_j}-\mathcal{G}_0 \tau_3 \partial_{k_i}\mathcal{G}_0 \frac{\partial g_z}{\partial k_j}\right]\nonumber\\
&=\frac{iq_i}{64\pi^3}\int{ d^3k} \,\text{tr}\left\{[\partial_{k_i}\mathcal{G}_0,\tau_3] \mathcal{G}_0 \frac{\partial g_z}{\partial k_j}\right\}\nonumber\\
&=\frac{-q_i}{16\pi^3}\sum_{\ell m}\int{ d^3 k} \frac{1}{(k_t^2-|g|^2+i\eta)^2}\epsilon_{\ell m} g_\ell \frac{\partial g_m}{\partial k_i}\frac{\partial g_z}{k_j}\nonumber\\
&=\frac{-iq_i}{32\pi^2}\sum_{\ell m}\int{ d {\mathbf k}} \frac{1}{|g|^3}\epsilon_{\ell m} g_\ell \frac{\partial g_m}{\partial k_i}\frac{\partial g_z}{\partial k_j}.
\end{align}
For the infinite system $p$-wave superfluid this results in (no summation convention)
\begin{eqnarray}
	\frac{i q_i \epsilon_{ij}}{32 m \pi^2}\int d{\mathbf{k}}\frac{v^2 k_j^2}{(\xi_{\mathbf{k}}^2+v^2 \mathbf{k}^2)^{3/2}}=\frac{i q_i \epsilon_{ij}}{16\pi}\kappa_a^{\infty}.
\end{eqnarray}

{\bf The coefficient $\boldsymbol{\kappa_b}$}. %%%%%%% \kappa_b
For convenience, we consider one of the correlators giving rise to the CS coefficient
\begin{multline}
\frac{i q_t}{64\pi^3}\int{ d^3k}\, \text{tr}[\partial_{k_t}\mathcal{G}_0 (\partial_{k_x}g\cdot\tau) \mathcal{G}_0 (\partial_{k_y}g\cdot \tau)\\
-\mathcal{G}_0 (\partial_{k_x}g\cdot\tau) \partial_{k_t}\mathcal{G}_0 (\partial_{k_y}g\cdot \tau)]\\
=\frac{iq_t}{32\pi^2}\int{d\mathbf{k}}\frac{\epsilon_{\mu\nu\lambda} g_\mu\partial_{k_x}g_\nu\partial_{k_y}g_\lambda}{|g|^3}.
\end{multline}

For the infinite system $p$-wave superfluid, we get
\begin{equation}
	\kappa^\infty_b=\frac{1}{4\pi}\int d\mathbf{k}\frac{\left(\frac{\mathbf{k}^2}{2m}+\epsilon_F\right)v^2}{\left[v^2 \mathbf{k}^2+\left(\frac{\mathbf{k}^2}{2m}-\epsilon_F\right)^2\right]^{3/2}}=\Theta(\epsilon_F).
\end{equation}

\subsection{The vanishing terms}

We provide an argument for the decoupling of the $a$ and $b$ fields, as well as for the vanishing of all mass terms for the field $b$.

{\bf The decoupling of the fields $\boldsymbol{a}$ and $\boldsymbol{b}$}. %%%%%%% No a.b
It can be shown that the integrand of the correlator describing the coupling between $a$ and $b$,
\begin{equation}
\int d^3 k\, \text{tr}\left[\mathcal{G}_0(k+\frac{q}{2})j_a^\mu(\mathbf{k})\mathcal{G}_0(k-\frac{q}{2})j_b^\nu(\mathbf{k})
\right],
\end{equation}
is always odd under $k\to -k$. Therefore, it vanishes to all orders in $q$ following an integration over $k$.\\

{\bf The absence of mass terms for the field $\boldsymbol{b}$}. %%%%%%% no b^2
In first order in the gradient expansion we find the following contribution to the mass of $b$:
\begin{equation}
\frac{-i}{2(2\pi)^3}\int{d^3k}\text{tr}\left[ \mathcal{G}_0 \left(-\frac{\tau_z}{2m}\right)\right]=-\frac{n}{2m}.
\label{eq:Contribution1ForbSquared}\end{equation}
Another contribution appears in second order (no summation convention),
\begin{multline}
 \frac{i}{32\pi^3}\int{d^3k}\text{ tr}\left(\mathcal{G}_0 j_b^i\mathcal{G}_0 j_b^i \right)=\\
 \frac{1}{16\pi^2}\int{d\mathbf{ k}}\frac{|g|^2(\partial_{k_i}g\cdot\partial_{k_i}g)-(g\cdot\partial_{k_i}g)^2}{|g|^3},
\end{multline} 
where in the infinite system $p$-wave superfluid we get, after integration over the angle of $\mathbf{k}$,
\begin{equation}
\frac{1}{16\pi}\int_0^{\Lambda_k}{d|{\mathbf k}|}\frac{v^2|{\mathbf k}|}{|g|^3}\left(\frac{{\mathbf k}^4}{2m^2}+2\epsilon_F^2+v^2{\mathbf k}^2 \right). 
\label{eq:Contribution2ForbSquared}
\end{equation}
While each is formally divergent, the sum of the two contributions, Eqs.~(\ref{eq:Contribution1ForbSquared}) and (\ref{eq:Contribution2ForbSquared}), now converges to zero,
\begin{multline} 
\lim\limits_{\Lambda_k\rightarrow\infty}\left[\frac{n}{2m}-\frac{1}{16\pi}\int{d|{\mathbf k}||{\mathbf k}|}\frac{v^2}{|g|^3}\left(\frac{{\mathbf k}^4}{2m^2}+2\epsilon_F^2+v^2{\mathbf k}^2 \right) \right]\\
=\lim\limits_{\Lambda_k\rightarrow\infty}\left(\frac{n}{2m}-\frac{\partial_{m}n}{4} \right)-\frac{|\epsilon_F|+m v^2}{8\pi}\kappa^\infty_a=0.
\end{multline}

\end{document}